\documentclass[aps,prl,preprint,groupedaddress, showkeys]{revtex4-1}
\usepackage{graphicx}
\usepackage{amsmath}
\usepackage{color}
\usepackage{lineno}
\usepackage{color,soul}

\begin{document}

\hyphenpenalty=1000

%\linenumbers

%\preprint{}


\title{Guided droplet transport on synthetic slippery surfaces inspired by a pitcher plant}



\author{Finn Box}
\affiliation{Mathematical Institute, University of Oxford, Woodstock Road, Oxford}

\author{Chris Thorogood}
\affiliation{Department of Plant Sciences, University of Oxford, South Parks Road, Oxford}

\author{Jian Hui Guan}
\email[]{jian.guan@eng.ox.ac.uk}
%\thanks{}
\affiliation{Department of Engineering Science, University of Oxford, Parks Road, Oxford}


\date{\today}

\begin{abstract}
We show how anisotropic, grooved features facilitate the trapping and directed transport of droplets on lubricated, liquid-shedding surfaces. Capillary action pins droplets to topographic surface features, enabling transport along the feature while inhibiting motion across (or detachment from) the feature. We demonstrate the robustness of this capillary-based mechanism for directed droplet transport on slippery surfaces by combining experiments on synthetic, lubricant-infused surfaces with observations on the natural trapping surface of a carnivorous pitcher plant. Controlling liquid navigation on synthetic surfaces promises to unlock significant potential in droplet-based technologies. Our observations also offer novel insight into the evolution of the \textit{Nepenthes} pitcher plant, indicating that the `pitfall' trapping mechanism is enhanced by the lubricant-infused, macroscopic grooves on the slippery peristome surface, which guide prey into the trap in a way that is more tightly controlled than previously considered.
\end{abstract}



\maketitle

%\section*{Introduction}
%One or two sentences providing a basic introduction to the field, comprehensible to a scientist in any discipline.
%

%%% Start of main Text %%%

%%% Introduction %%%

\section{Introduction}
\begin{figure}[h]
\includegraphics[width=0.5\textwidth]{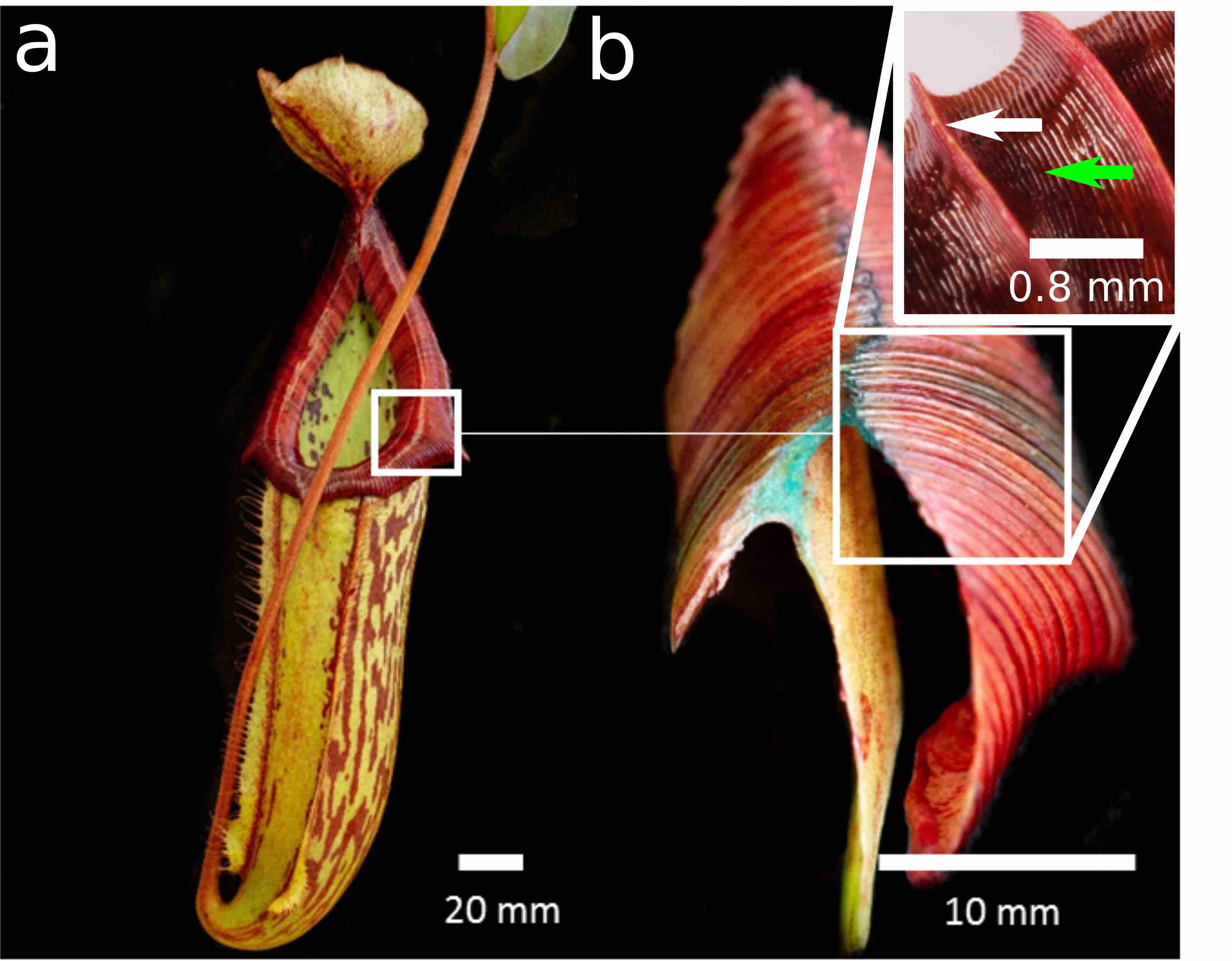}
\caption{\textbf{The \textit{Nepenthes} peristome}. a. The trap of a \textit{Nepenthes} pitcher plant (\textit{N}. x \textit{mixta}). b. A portion of the peristome showing both the inward (left) and outward (right) slanting regions. Inset: light micrograph showing macroscopic grooves formed by macroscopic ridges (white arrow) and microscopic grooves (green arrow).}
\label{figure1}
\end{figure}

Living organisms have evolved a fascinating array of functional surfaces for interacting with their environments \citep{Barthlott2017}. Our understanding of how liquids can be manipulated has been transformed by natural surfaces, such as those of the water-repellent lotus leaves (\textit{Nelumbo nucifera}) \citep{Barthlott1997}, the water-collecting wing-cases of desert beetles (\textit{Stenocara} spp.) \citep{parker2001water} and the water-removing skin of the box-patterned gecko (\textit{Lucasium steindachneri}) \citep{Watson20141396}. Recently, the wettable surface of the carnivorous pitcher plant \textit{Nepenthes} (Figure \ref{figure1}) has inspired new types of lubricant-infused, slippery surfaces \citep{wong2011bioinspired, JDsmith_slips}. Pitcher plants produce specialised leaf-derived `pitfall' traps to attract, capture, retain, kill and digest animal prey to enable them to survive in nutrient-poor environments \citep{Thorogood2018, moran_pitcherplant}. A key trapping feature is the peristome, which has sloping, macroscopic ridges (Figure \ref{figure1}a), in turn made up of microscopic ridges (Figure \ref{figure1}b) \citep{chen2016continuous}. The peristome is slippery when wet, by virtue of a lubricating film of water which condenses on the fully wettable surface, leading insects to aquaplane into the trap \citep{Bauer2008}. 

\begin{figure*}[ht]
\includegraphics[width=\textwidth]{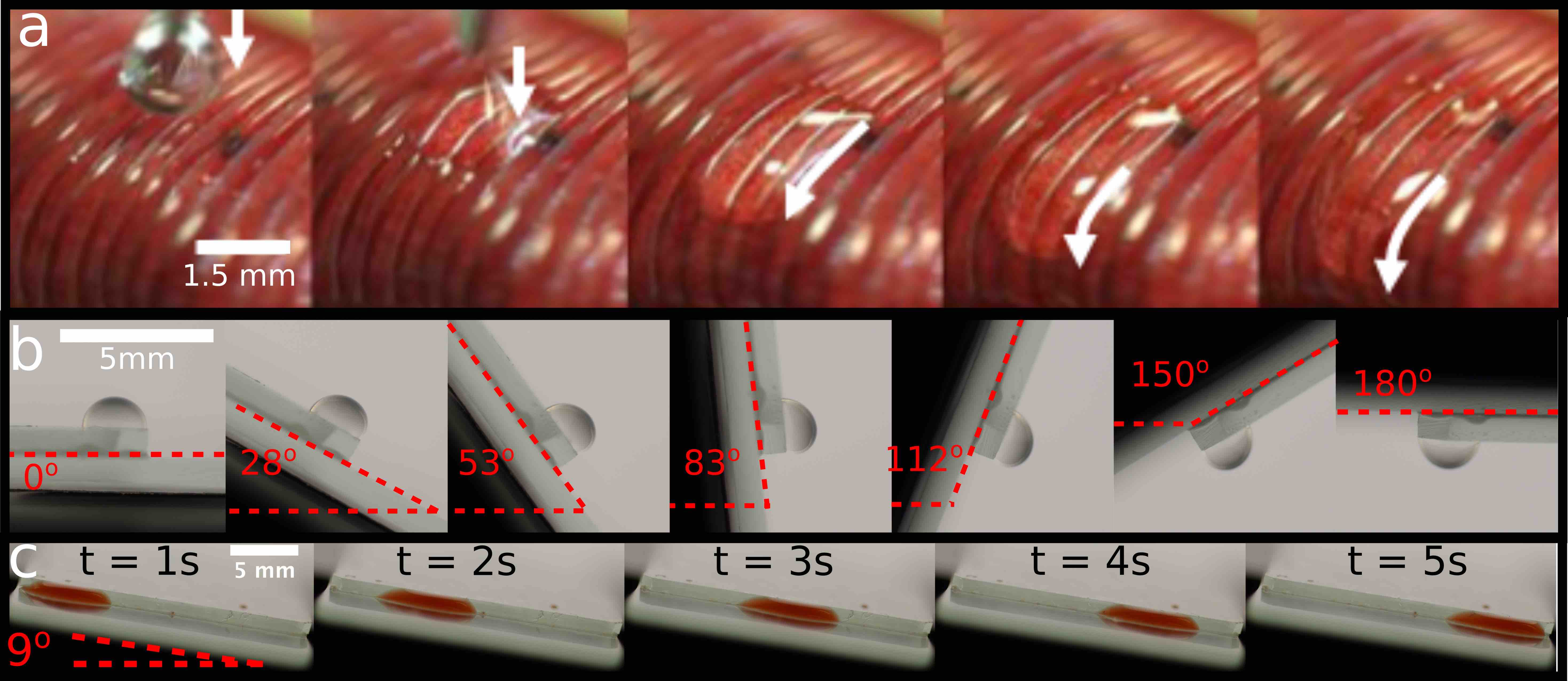}
\caption{\textbf{Droplet motion on both the surface of the \textit{Nepenthes} peristome and artificial surfaces.} a. Time sequence of a drop of rapeseed oil placed on the peristome moving along the macroscopic grooves (white arrow), rather than across them. b. A 5 $\mu \rm L$ droplet placed above a lubricated step ($h = 1.1 \ \rm mm$) remains pinned at the edge of the step as it is rotated from 0 to 180$^\circ$, where the axis of rotation is parallel to the step. c. Time sequence of a dyed water droplet attached to the same lubricated step sliding along the step at a tilt angle of 9$^\circ$, where the axis of rotation is perpendicular to the step.}
\label{figure2}
\end{figure*}

Lubricated surfaces in a technological context are referred to as Slippery Liquid-Infused Porous Surfaces (SLIPS) \citep{wong2011bioinspired} and, equivalently, Lubricant-Impregnated Surfaces (LIS) \citep{JDsmith_slips}. Both SLIPS and LIS are modelled on the peristome surface of the \textit{Nepenthes} trap in that they are preferentially wetted by a film of lubricating liquid that repels foreign liquids (immiscible to the lubricating liquid) \citep{wong2011bioinspired, JDsmith_slips}. Since their conception, these surfaces have often out-performed their conventional superhydrophobic counterparts in terms of their liquid-shedding abilities \citep{wong2011bioinspired, JDsmith_slips, guan2015evaporation, keiser2017dropfriction, guan2017drop}. However, the lack of drop-solid interaction on these surfaces also means that controlling the motion of liquid droplets is inherently difficult \citep{guan2015evaporation, daniel2017oleoplaning}. The lack of controlled droplet transport limits the application of these liquid-shedding surfaces in droplet-based technologies. Identifying new mechanisms for harnessing the directional mobility of droplets will inform the design of synthetic surfaces that transport droplets in a controlled way. Such mechanisms could be applied directly to droplet-based technologies such as rainwater harvesting \citep{dai2018} and anti-fogging coatings \citep{howarter2008self}, as well as to rapidly expanding new technologies such as Micro-Electro-Mechanical Systems (MEMS) \citep{ashraf2011micro} and digital microfluidic devices \citep{choi2012digital}.
Examining functional natural surfaces in vivo may also offer insights into the evolution of natural systems \citep{shirtcliffe2009learning}. 
Whilst the aquaplaning trapping mechanism of the carnivorous \textit{Nepenthes} pitcher is well documented \citep{Thorogood2018, Bohn14138, chen2017novel}, the functionality of the grooves on the peristome surface remains relatively unexplored. Here, we show that capillary action pins droplets to topographical features (grooves), enabling motion along the direction of the feature whilst inhibiting motion across the feature. In doing so, we decipher the mechanism by which prey is guided into the pitcher plant trap, and also provide a means of generating pathways that can control and guide the motion of droplets on slippery, synthetic surfaces.

%%% End of introduction %%%
\definecolor{marker1}{RGB}{62,38,168}
\definecolor{marker2}{RGB}{52,122,253}
\definecolor{marker3}{RGB}{18,190,185}
\definecolor{marker4}{RGB}{200,193,41}

%%% Results %%% 

\section{Results}

\begin{figure*}[ht]
\includegraphics[width=\textwidth]{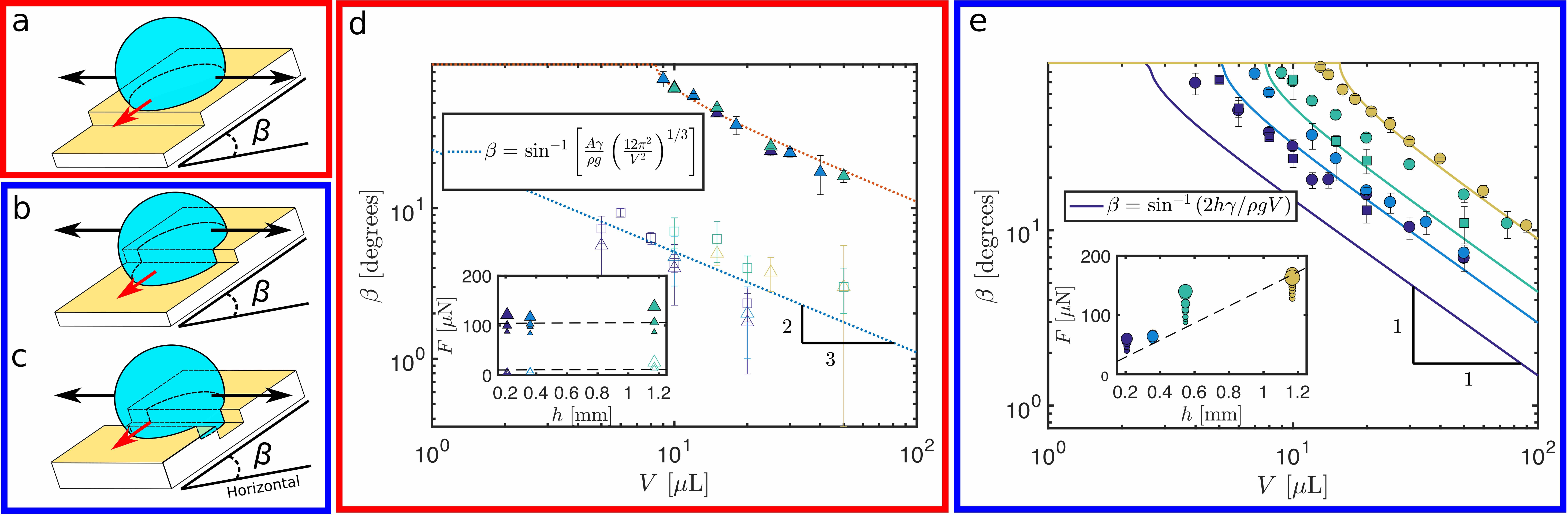}%
\caption[width=\textwidth]{\textbf{Schematics of model geometries and theoretical predictions.} a. A droplet above a step. b. A droplet attached to a step from below. c. A droplet in a trench (the black arrows indicate the possible motions of the droplets along the topographical features and red arrows indicate the direction droplets detach for inclination angles larger than $\beta$). (a-c illustrate the experimental set up for all model geometries, all faces of the non-planar surfaces were rendered as SLIPS; the thin lubricant layer is represented in yellow and water droplets in blue. Here we do not depict the lubricant meniscus, which has been considered elsewhere \citep{orme2019droplet}.) d. The inclination angle $\beta$ at which a droplet of volume $V$ detaches from above a step orientated parallel to the axis of rotation (filled markers) is an order of magnitude larger than the inclination required for a drop to slide along the step (empty triangles) or along a trench (empty squares). The dotted lines represent least-squares fits of the radius-dependent scaling (\ref{eq:above}) with prefactor $A = 0.130\pm0.004$ (detaching) and $A = 0.0131\pm0.002$ (sliding).  Inset: The force (weight) required to overcome capillary pinning of the drop is independent of step height. e. The inclination angle $\beta$ at which a droplet of volume $V$ detaches from below a step (circles) or a trench (squares). The solid lines represent the scaling (\ref{eq:below}) for different topography heights. Inset: The force (weight) required to overcome capillary pinning of the drop increases with feature height; the influence of the effective drop radius is secondary (marker size indicates drop volume). In d-e, the height of the elements are colour coded such that $h = 0.21$\,mm (\textcolor{marker1}{navy}), $0.36$\,mm (\textcolor{marker2}{blue}), $0.55$\,mm (\textcolor{marker3}{green}), $1.17$\,mm (\textcolor{marker4}{yellow}).}
\label{dataplots}
\end{figure*}

\textit{\textbf{Non-arbitrary droplet transport on lubricated topographies}} - The directional liquid transport observed along the macroscopic grooves present on the \textit{Nepenthes} peristome (Figure \ref{figure2}a) can be reproduced on artificial lubricated surfaces. Using variations in surface topography, we fabricated lubricated surfaces capable of trapping, retaining and directing the transport of liquid droplets (Figure \ref{figure2}b-c). We used two non-planar model geometries: a step (with a droplet either attached to the step or situated above the step) and a single trench (Figure \ref{dataplots}a-c). All faces of the non-planar surfaces were rendered as SLIPS such that they were infused entirely with a lubricating liquid to ensure that all faces bore the same low-hysteresis characteristics as those seen on planar SLIPS.
Guan \textit{et al}. previously reported the formation of a meniscus due to the distortion to the lubricant layer by a step when the lubricant thickness is comparable to the step height \citep{guan2017drop}. Here, we assume the shape of the lubricating liquid roughly follows the shape of the macroscopic feature, since the thickness of the lubricating layer is orders of magnitude smaller than the size of the feature. 
%
%It has been previously reported that the lubricating liquid tends to collect in the inner corner of a step and a meniscus can form to a certain lateral length, $L$, which depends on the ratio between the lubricant thickness and the height of the step. $L$ was measured to be $0.2 \ \rm mm$ for a $50 \ \mu \rm m$ tall step and a lubricant thickness of $13 \ \mu \rm m$ {\citep{guan2017drop}}. In our work, we assume the shape of the lubricating liquid roughly follows the macroscopic surface, having a lubricant thickness orders of magnitude smaller than the size of the features.}
%
We found that droplets in contact with a non-planar feature experience a strong capillary adhesion (pinning) that inhibits detachment, but are relatively free to slide along the feature in the direction where the topographical variation was minimal or zero (represented by the black arrows in Figure \ref{dataplots}a-c). The non-planar feature can therefore be exploited to trap and retain droplets – even when held upside-down (Figure \ref{figure2}b) – or to control the direction of liquid transport across the slippery substrate (Figure \ref{figure2}c). When sliding along a feature, the droplets exhibited negligible contact angle hysteresis, which is comparable to the behaviour of liquids moving along the grooves of the \textit{Nepenthes} peristome and on planar SLIPS – for which droplets slide at remarkably shallow angles, typically a few degrees \citep{wong2011bioinspired, JDsmith_slips, guan2015evaporation, keiser2017dropfriction, guan2017drop, daniel2017oleoplaning, schellen_direct_observation}

We consider the free energy of the system to understand better the directional fluid motion on artificial surfaces with variations in topography. Mechanical equilibrium requires that the free energy of a system must be a minimum. In the case of a liquid drop, this implies that the pressure, imposed by the curvature of the exposed interface, must be uniform otherwise it would give rise to internal flow \citep{de2013capillarity}. Furthermore, its local contact angle must satisfy the Young-Dupr\'e relation: this forces droplets to acquire unique shapes when in contact with solids (the shapes depend on the topography of the solid). Due to symmetry, translational motion of a droplet along topographical features can be produced with a vanishingly small alteration in its shape.  By contrast, the symmetry of the droplet across the grooves is not continuous, and therefore a non-negligible deformation of the droplet is required to produce a continuous translation \citep{wells2018snap}. An additional energy must be spent, or equivalently, a threshold force must be surpassed. Variations in the topography of the surface therefore have direct influence on the energy landscape and imposes energy barriers over which liquid droplets must overcome if motion is to occur \citep{huhenergybarrier}. 

We performed experiments on lubricated surfaces to investigate the retaining influence that a topographic feature exerts on a liquid droplet. We carefully positioned droplets either above or below (and in contact with) various topographic features and measured the angle of repose, $\beta$. We define $\beta$ as the critical angle at which the weight of the drop becomes large enough to overcome the capillary adhesion pinning the droplet to the feature, such that the droplets either detach or fall from the obstacle or slide along the feature depending on the orientation of the axis of rotation. The results are shown in Figure \ref{dataplots}d-e for the various geometries shown in Figure \ref{dataplots}a-c.

For a drop positioned above a step (see Figure \ref{dataplots}a), an inclination almost an order of magnitude larger is required to detach the drop (in a direction perpendicular to the length of the feature) than to move along the top of the step (in the direction parallel to the feature). In both cases, however, the angle of repose scales with droplet volume as $\beta \sim V^{-2/3}$. Pinning of the drop to the top of the step results from interfacial tension acting on the contact line between the droplet and the lubricating fluid. This capillary adhesion results from the interfacial line force, $F_c \sim 2\pi R \gamma$, where $R=(3V/2\pi)^{1/3}$ is the effective radius of a hemispherical drop. From a vertical force balance between $F_{c}$ and the weight of the drop, $F_w = \rho g V \sin\beta$, we find that the angle of inclination required to overcome pinning scales as,
\begin{equation}
\beta \sim \sin^{-1} \left[\frac{\gamma }{\rho g}\left(\frac{12 \pi^2}{V^2}\right)^{1/3}\right].
\label{eq:above}
\end{equation}
We consider that only a fraction of the line force opposes gravity relative to the total line force $F_c =  2A\pi R\gamma$, where A is a constant which gives the relative fraction of the contact line that contributes to pinning. From fits to the data (dotted lines in Figure \ref{dataplots}d) we determine the prefactor $A$ and estimate that pinning results from $\sim 13\%$ of the contact line when motion occurs across the feature (i.e. in detachment), whereas only $\sim 1\%$ of the total line force acts against gravity when the droplet slides along the feature. The force (weight) required to disturb a drop from its pinned position clearly demonstrates that the relative strength of the capillary adhesion perpendicular to a feature is an order of magnitude larger than along the direction of the topography (see inset to Figure \ref{dataplots}d), and this anisotropy clearly enables guided motion of droplets over slippery surfaces. 

\begin{figure*}[t]
\includegraphics[width=\textwidth]{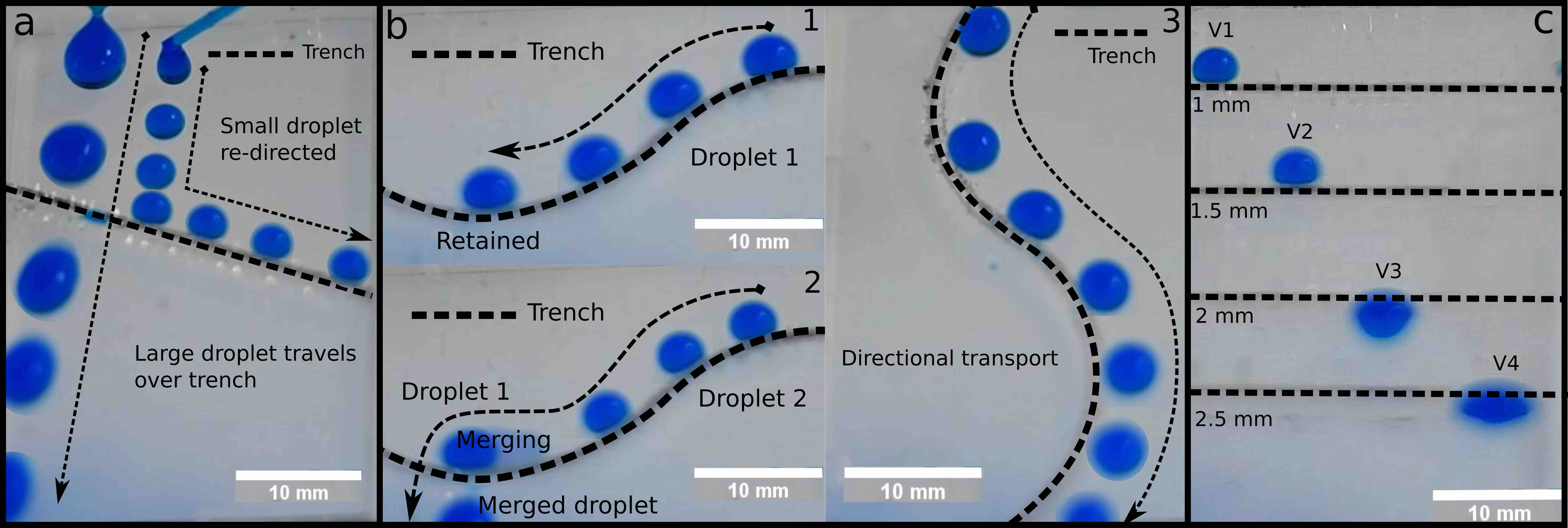}%
\caption[width=\textwidth]{\textbf{Composite images of guided droplet transport.} a. A SLIPS with a diagonal trench (width, $w = \rm 1 \ mm$ and depth, $h = \rm 1\ mm$). A small droplet ($V\approx50\ \mu \rm L$) travelling at an initial velocity of $1.1 \ \rm mms^{-1}$ is  re-directed by the trench  whereas a larger droplet ($V\approx125\ \mu \rm L$) travelling $3.2 \ \rm mms^{-1}$ is not. b. A $50 \ \mu \rm L$ droplet travelling along and above a curved trench is retained in a local minima (1) until it merges with a second $50 \ \mu \rm L$ droplet and travels over the trench ($w = \rm 1\ mm$)(2). The gravity-driven transport of a $50 \ \mu \rm L$ droplet is dictated by the presence of the trench (3). c. Droplets of different volumes, $v$, are retained by trenches of different widths, $w$. $v\rightarrow w$: $50\ \mu \rm L \rightarrow 1 \ \rm mm$, $80\ \mu \rm L \rightarrow 1.5 \ \rm mm$, $150\ \mu  \rm L \rightarrow 2 \ \rm mm$, $300\ \mu  \rm L \rightarrow 2.5 \ \rm mm$. All surfaces were inclined at 45$^\circ$.}
\label{differentsurfaces}
\end{figure*}

Similar to the case of a drop above a step, anisotropic mobility occurs for drops attached to the underside of a step or in a trench/groove (Figure \ref{dataplots}b and \ref{dataplots}c, respectively); when sliding along a feature the angle of repose is dependent on the droplet size (see trench data in Figure \ref{dataplots}d). However, the inclination angle required for detachment instead is found to be inversely proportional to the drop volume, $\beta\sim V^{-1}$ (see Figure \ref{dataplots}e). The force pinning the drop to the feature increases with the height of the feature (shown in the inset to Figure \ref{dataplots}e) in accord with ref. \citep{orme2019droplet} which reported that the capillary force of a $2\,\mu$L drop scales linearly with the height of a step. Noting that the influence of the droplet radius is secondary to the feature height, we balance the weight of the drop, $\rho g V \sin \beta$, with the line force exerted by the droplet contact line attached to the feature, $2 h \gamma$, where $h$ is the height of the feature, and find, 
\begin{equation}
\beta \sim \sin^{-1} \left(\frac{2h\gamma}{\rho g V}\right).
\label{eq:below}
\end{equation}
The agreement between predictions (with no fitting parameters) and experimental data (solid lines in Figure \ref{dataplots}e) confirms that the pinning arises predominantly from the portion of the drop in contact with the part of the feature that is orthogonal to the plane of the surface. 

Although the surfaces are completely slippery, the pinning mechanism can be understood in terms of Gibb’s inequality, which reads\citep{Kalinin2009, herminghaus2008wetting},
\begin{equation}
\theta_{\rm e} \leq \theta \leq (180^\circ - \phi) + \theta_{\rm e}
\end{equation}
\noindent where $\theta_e$ is the equilibrium contact angle of the drop. As a droplet approaches the sharp edge of a topographic feature, of angle $\phi$, the contact line remains pinned until the contact angle of the droplet $\theta$ reaches a value of $\theta = \theta_{\rm e} + (180 -\phi)$, after which the contact line advances \citep{Kalinin2009, herminghaus2008wetting}. (In the case wherein the trailing edge leaves a corner, a condition of $\theta = \theta_{\rm e} - (180 -\phi)$ should be satisfied). Further analysis is required to understand the subtle contribution of the interface geometry, and the role of the `skirt' of lubricating liquid \citep{semprebon}, to the capillary pinning. 
A first step has been made towards this by Orme \textit{et al}. who recently showed attractive/repulsive interactions between the droplet's wetting `skirt' and the distorted oil layer due to the presence of a small step, depending on the droplet's initial position \citep{orme2019droplet}. In particular,  Orme \textit{et al}. observed that a drop positioned above a step sits from distance from the step edge, whereas a drop positioned below a step attaches to the non-planar portion of the feature and overlaps the step (providing the drop is larger than the size of the step).
%
%\hl{They find that the strength of this interaction scales with the ratio between the thickness of the lubricant and the step height \mbox{\citep{orme2019droplet}}.}

%{\citep{orme2019droplet}}

\section{Discussion}

Our experiments confirm that droplets adhere to topography \textit{via} interfacial tension, and preferentially travel along topographic features rather than across them. Our results also suggest that, for a given $\beta$, lubricated topographies are capable of capturing droplets up to a critical volume, $V$. We performed experiments to test this hypothesis and demonstrated that droplets moving along a planar SLIPS can be re-directed (Figure \ref{differentsurfaces}a, see Movie S1) or captured (Figure \ref{differentsurfaces}c, see Movie S4) by a topographic feature. This finding exposes a potential mechanism for developing systems in which the transport of droplets is guided by these topographically induced `energy railings'. In the absence of any pinning forces, a droplet on a lubricated surface is able to survey freely the global energy landscape, and would therefore follow the most ‘energy-saving’ path. The relationship between  $\beta$ and $V$ also allows surfaces to be fine-tuned to capture and direct droplets of specific sizes as they travel across slippery surfaces (Figure \ref{differentsurfaces}a-c). Using this rationale, we created surfaces with pre-determined variations in topography to demonstrate the non-arbitrary transport of highly mobile droplets (Figure \ref{differentsurfaces}b, see Movie S2 and 3).

\begin{figure*}[t]
\includegraphics[width=\textwidth]{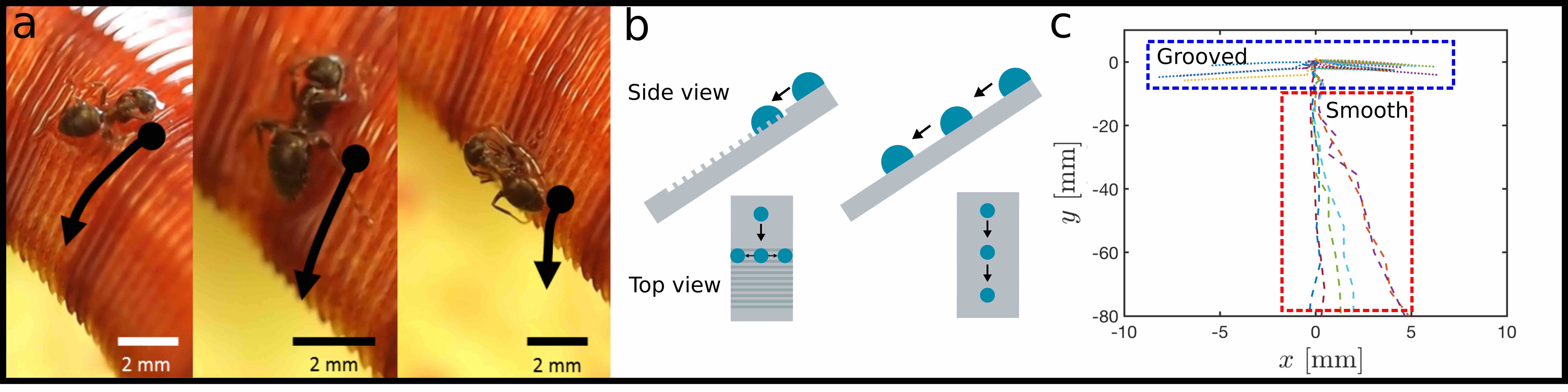}%
\caption{\textbf{Non-arbitrary transport of droplets or prey on lubricated topographies.} a. An ant slides into the pitcher by slipping along the grooves. b. Schematics showing top and side view of drop transport along a grooved cylindrical surface compared to a smooth one. c.  Droplet trajectories on grooved (dots, blue box) and smooth (dashes, red box), curved SLIPS (All surfaces were inclined at 45$^\circ$).}
\label{insect figure}
\end{figure*}

Although guided droplet transport has also been demonstrated on conventional superhydrophobic tracks, it is more difficult to achieve due to the inherent contact angle hysteresis which must be overcome to produce motion \citep{mertaniemi2011superhydrophobic, Yunusa_directional_super, xu2008directing}. Without external forces, energy barriers can only be overcome with an excess of potential energy stored within the drop, whether that be through an increasing tilt angle of the surface or an increasing volume. In the case of superhydrophobic surfaces, the effect of such changes in the momentum of a drop would render it uncontrollable. Lubricated topographies, however, would be able to overcome such shortcomings thanks to the pinning introduced by the feature and the droplet's tendency to adhere to the lubricating liquid (whilst being completely mobile). Furthermore, directed motion on lubricated topographies could be created with a vanishingly small tilt of the surface \citep{wong2011bioinspired, JDsmith_slips, guan2015evaporation, keiser2017dropfriction, guan2017drop, daniel2017oleoplaning, schellen_direct_observation}.

\textbf{\textit{Droplet transport on the} Nepenthes \textit{peristome}} - Pitfall trap production in carnivorous plants is costly, so strong selection pressures should act on traps to maximise their prey intake \citep{Thorogood2018, moran1998foliar}. Therefore, the ornate, non-planar features on the pitcher peristome are presumably functional and associated with a strong selective advantage. In nature, insects aquaplane off the inward-facing portion of the peristome into the trap interior (Figure \ref{figure1}a-b) \citep{Bohn14138}. We explored the function of non-planar features by observing both foreign droplets, and prey on the \textit{Nepenthes} peristome. Foreign liquid droplets placed on either the inward or outward slanting regions of the peristome migrated towards the pitcher’s interior or exterior, respectively, by preferentially travelling along the macroscopic grooves, consistent with our experiments on synthetic surfaces. Ants (the typical prey of \textit{Nepenthes} pitcher plants) were also observed to slide on the wetted surface of the peristome. The tarsi of the ants repeatedly slid within the film of water retained along the grooves, driving insects into the trap (Figure \ref{insect figure}a), a mechanism described in greater detail by other authors \citep{chen2016continuous, Bauer2008, Bohn14138, chen2017novel}. Based on our characterisation of non-arbitrary droplet transport on lubricated topographies, we propose that mechanically, the ease of droplet or prey migration along the grooves can be attributed to the natural slanting of the peristome surface combined with the negligible friction of the slippery peristome surface. The surface is highly wettable, so it is energetically favourable for water to spread onto, and be retained by the grooves; forming a stable thin film on the peristome surface \citep{chen2016continuous}. This removes direct contact between the foreign liquid droplet and the peristome surface, and thus the associated contact angle hysteresis caused by contact line pinning. The macroscopic grooves may therefore guide prey into the trap, under the influence of gravity, as motion across the grooves is energetically unfavourable.

Finally, to test preferential transport as a functionality of the grooves, we fabricated cylindrical surfaces (of curvature $\kappa = 0.05 \ \rm m^{-1}$) both with grooves aligned orthogonal to the centreline (the grooves were 0.5 mm tall, 1 mm wide with a 1.5 mm gap between them). The curved surfaces were positioned at $\approx 45^\circ$, replicating the natural orientation of the peristome. Droplets were released onto the centre line of the surface. When a $\sim10\,\mu$L droplet, travelling down a surface (under the influence of gravity) at $\sim1$\,mm\,s$^{-1}$, came into contact with a groove oriented orthogonal to the initial liquid transport direction, motion of the droplet in its original direction of travel was halted and the droplet instead travelled along the grooves (Figure \ref{insect figure}b-c). In the absence of grooves, however, water droplets displayed preferential motion along the length of the surface (Figure \ref{insect figure}b-c). These results support our hypothesis that grooves facilitate controlled, non-arbitrary transport of prey along the contoured, lubricated peristome surface.\raggedbottom

\section{Conclusion}

In summary, our data show that capillary pinning to variations in surface topography facilitates the trapping and retaining of droplets of specific sizes on otherwise completely slippery surfaces. 
Anisotropic features such as grooves, with zero variation in surface topography in one direction, can therefore be used to guide the motion of liquid droplets, since motion is impeded in the direction of maximal variation in surface topography. 
Our observations of foreign liquid droplets on both artificial, lubricant-infused surfaces and on natural surfaces of a tropical pitcher plant, together demonstrate that transport is robust, non-arbitrary and favours translation along slippery grooves, rather than across them. We suggest that this may play a crucial role in the `pitfall' trapping mechanism of this carnivorous plant, by directionally driving prey into the trap in a way that is more tightly controlled than previously understood. This offers novel insights into the role of the peristome in the evolution of the \textit{Nepenthes} trapping mechanism. 
The findings also provide a biomimetic means of transporting and sorting droplets that is straightforward to implement in droplet-based fluidic devices. 
The use of SLIPS/LIS in droplet-based fluidic devices has been hindered by an inability to control droplet motion, and has been attributed to a lack of droplet-surface contact. We have demonstrated that, {\textit{via}} inclusion of topographic features, control of droplets on SLIPS/LIS can be achieved without direct contact between droplet and surface. This provides a means of designing slippery surfaces that can guide droplets in a controlled manner, which could have profound implications for enhancing the functionality of droplet-based technologies.
For example, incorporating droplet-guiding railings into the surface of solar panels could increase the efficiency of hybrid devices, that harvest energy from both sunlight and rain, by guiding raindrops onto functional regions of the surface that are populated with triboelectric nanogenerators \citep{liu2018integrating}.

\section{Materials and Methods}

\textbf{\textit{Nepenthes}} - Plants (\textit{Nepenthes} x \textit{mixta}) were obtained from the University of Oxford Botanic Garden, selected for their broad peristomes which were easy to manipulate, and typical of the genus, and for their numerous pitchers (for repeatability). For the liquid transport experiments, the peristome surface of the plant was initially wetted with a light spray of double-distilled water.

\textbf{Liquid transport on \textit{Nepenthes} peristome} - We used time-lapse photography to capture the motion of liquids which are immiscible to the lubricating water film. We first deposited small droplets (approx. $10 \mu$L) of rapeseed oil on both the inward and outward slanting regions of the peristome with a syringe. Portions of the peristome with the maximum slanting and width were selected (see Figure \ref{figure1}).

\textbf{Lubricated topographies} - We created the two model geometries used in this study by bonding a number of thin glass coverslips of known thickness into a composite with variations in surface topography. The height of the topography was measured using an optical microscope. To construct the surface with a single trench, we used the same method and placed two stacks of coverslips a known distance apart (Trench 1: $w$ = 0.7 mm, $h$ = 0.21 mm, Trench 2: $w$ = 1.2 mm, $h$ = 0.55 mm). To create more complex surfaces, i.e. surfaces with multiple trenches and lubricated tracks, we engraved clear Perspex using a laser cutter to achieve a depth of $\approx 2\rm mm$. The surfaces were rendered superhydrophobic with a nano-particle based coating (Glaco Mirror Coat Soft 99 Co.). They were subsequently immersed in a bath of silicone oil (Sigma-Aldrich, 20cSt) and withdrawn vertically at a controlled rate of $1 \ \rm mm s^{-1}$ to create a uniform lubricant layer of thickness, $t \approx 13 \ \mu \rm m$ \citep{guan2017drop}.

\textbf{Sliding angles and angles of repose}
Sliding angle measurements for water droplets on lubricated surfaces were carried out using a bespoke rotatable stage which was levelled using a spirit level prior to any measurements. A single droplet of water was dispensed onto the surface in contact with the non-planar feature. The surface was then rotated until the droplet started to move. The angle of repose was measured for water droplets in contact with the model geometries shown in Figure \ref{dataplots}a-c, for the axis of rotation both parallel and perpendicular to the non-planar feature. 

%%% End of results %%% 

%%% End of main text %%%

\section{Author Contributions}
%J.H.G. and F.B. conceived the research. F.B. and J.H.G. designed the experiments on the lubricated surfaces and F.B. carried out the experiments; C.T and J.H.G performed the \textit{in vivo} experiments. F.B. and J.H.G analysed and interpreted the data.  F.B., C.T. and J.H.G wrote the paper.
J.H.G. and F.B. designed the research. F.B., C.T. and J.H.G performed the research and authored the paper.

\section{Acknowledgments}
John Baker provided light micrographs of the \textit{Nepenthes} peristome; Sam Ibbott took video footage and photographs of insects on the \textit{Nepenthes} peristome. We thank Nicasio Geraldi and \'Elfego Ruiz-Guti\'errez for useful discussions. 

\section{Competing Interests} 
The authors declare that they have no competing financial interests.

\section{Data and materials availability}
All data needed to evaluate the conclusions in the paper are present in the paper and/or the Supplementary Materials. Additional data related to this paper may be requested from the authors.

\section{Correspondence}Correspondence and requests for materials should be addressed to J.H.G \\ (Email: jian.guan@eng.ox.ac.uk).

\bibliography{ref_abbrev}

\end{document}